# PIDS: A Behavioral Framework for Analysis and Detection of Network Printer Attacks

**Asaf Hecht**
Ben-Gurion University of the Negev
Beer-Sheva, Israel
hechta@post.bgu.ac.il

**Adi Sagi**
Ben-Gurion University of the Negev
Beer-Sheva, Israel
sagiad@post.bgu.ac.il

**Yuval Elovici**
Ben-Gurion University of the Negev
Beer-Sheva, Israel
elovici@bgu.ac.il

**Abstract.** Nowadays, every organization might be attacked through its network printers. The malicious exploitation of printing protocols is a dangerous and underestimated threat against every printer today, as highlighted by recent published researches. This article presents PIDS (Printers' IDS), an intrusion detection system for detecting attacks on printing protocols. PIDS continuously captures various features and events obtained from traffic produced by printing protocols in order to detect attacks. As part of this research we conducted thousands of automatic and manual printing protocol attacks on various printers and recorded thousands of the printers' benign network sessions. Then we applied various supervised machine learning (ML) algorithms to classify the collected data as normal (benign) or abnormal (malicious). We evaluated several detection algorithms, feature selection methods, and the features needed in order to obtain the best detection results for protocol traffic of printers. Our empirical results suggest that the proposed framework is effective in detecting printing protocol attacks, providing an accuracy of 99.9 with negligible fall-positive rate.

## 1 Introduction

Printers are part of every commercial and personal network today and therefore every network is at high risk of being attacked. In particular, network printers and multi-function printers (MFPs) have become attractive for cyber-attacks in many networks as they must expose many interfaces (both wired and wireless) and support many protocols in order to serve a broad base of in-house workers and ad hoc visitors. Previous researches and publications have demonstrated the feasibility of compromising all kinds of printers [9-13], more on those researches will be provided in the related work section.

Printing protocol attacks comprise a variety of malicious actions that an attacker can perform against the target printer, including among others: denial-of-service (DoS) attacks, privilege escalation, leaking print jobs or system files; and even initiating code execution on the printer itself [1].

In recent years, the threat against printers has received an increasing amount of attention from academia and the media [2-4] following the development of open-source attacking tool PRET: PRinter Exploitation Toolkit [5-8]. PRET is an extremely popular hacking tool dedicated to attacking network printers. Considering the aforementioned risks, the prevalence of networked printers, and the variety of protocols they use, our aim in this research is to evaluate whether supervised machine learning (ML) is an effective method for intercepting the ever-increasing attacks on printing protocols.

The main contributions of this study are as follows:

(1) Develop and assess a novel IDS framework for detecting attacks on printing protocols by benchmarking the effectiveness of various supervised ML algorithms methods that are trained on benign and malicious traffic.

(2) Create a first of its kind robust dataset of printing protocols behavior using a vast collection of malicious and benign traffic sessions. The dataset will be available for future research. [A link to the dataset will be added after the possible acceptance confirmation].

(3) Enhance PRET [5] with automatic mechanisms for executing multiple attack commands on multiple printers automatically.

(4) Provide the pillars for future work on securing other peripherical devices and daily office objects such as: 3D printers, smart IP phones, IP projectors, etc.

The rest of the paper is structured as follows. It first describes the related work (section 2) and then continues with presentation of the different components of our proposed PIIDS detection framework including the data collection phase, the detection model, training phase and the final evaluation phase (section 3).

## 2 Related Work

Several articles deal with the malicious usage of printing protocols such as PJL (Printer Job Language) and page description languages such as PostScript and PCL (Printer Command Language). In [9] a proof of concept demonstrated malicious access to the printer's file system using PJL. Other research on malicious PJL commands was conducted in [10]. The potential malicious use of PostScript was discussed by [11]. These three studies focusing on offensive tactics and other means of attacking printers were described thoroughly in a recent research [12] and can be executed easily with the research's attacking tool - PRET.

Other research on attacking printers [13] showed how firmware updates can be exploited to inject malicious firmware modifications into vulnerable embedded devices. This research focused on a case study on compromising RFU feature (remote firmware update) in HP LaserJet printers. In this study the researchers discovered a modification vulnerability which allows the arbitrary injection of malware into the printer's firmware via standard printed documents.

Regarding the use of ML for detecting attacks on printing protocols used by network printers, there has been no previous ML research. The closest research we found was in the field of securing additive manufacturing (also known as 3D printing). Research conducted by [14], focused on securing 3D manufactures printers and CNC milling machines. This research evaluated ML algorithms for the detection of cyber-physical attacks on the manufacturing devices. They suggested a system for protecting the manufacturing machines from malicious abnormal transmissions and attacks. The fundamental difference between our research and theirs is the fact that 3D printers and regular network printers are using totally different printing protocols.

Another study [15] employed ML algorithms to steal printed models (intellectual property) through side-channel attacks via noise and magnetic radiation of 3D printers. They, however, did not deal with attacks on conventional network printers.

To date, most of the research conducted on securing network printers and MFPs have focused on the printers' threats and vulnerabilities [16] and how to exploit them [12]. Research [17] has also proposed a method used to securely install and operate printers in the organization.

To the best of our knowledge, no previous research has focused on the detection of attacks on conventional (non-3D) printer protocols by training and testing supervised ML classifiers on a collection of malicious and benign traffic. PIDS is tailored to detect protocol attacks on networked printers and MFPs. For that purpose, we created a new dataset with thousands of automatic and manual printing protocol attacks on various printers, alongside with thousands of real-world recorded printers' benign traffic. In addition, our research presents a different and robust feature extraction technique. In the following sections we present our proposed PIDS system, experiments we conducted in order to evaluate its performance, and discuss our evaluation results.

## 3 Proposed PIDS System

In this paper, we suggest an IDS that employs supervised ML to discover and subsequently issue alerts when malicious protocol traffic is detected. Figure 1 presents the creation process of the dataset and the evaluation phase of different classifier algorithms. Figure 2 describes the proposed PIDS architecture in the network.

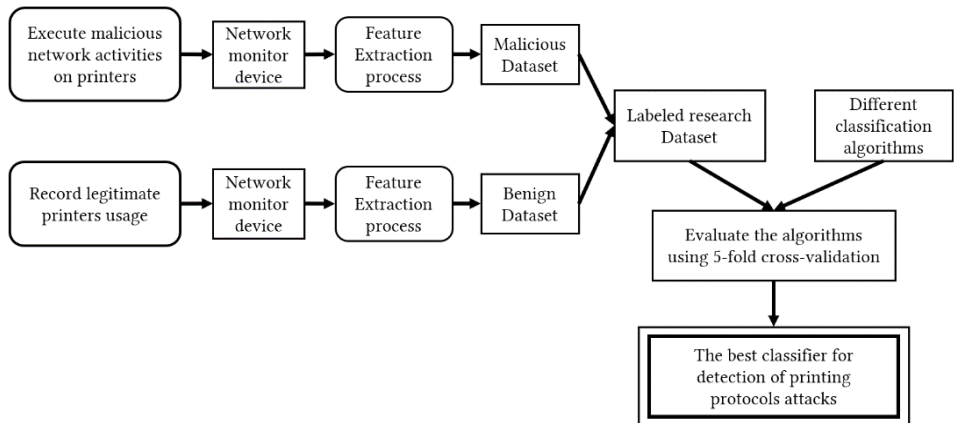

**Fig. 1.** Training and evaluating the best detection classifier

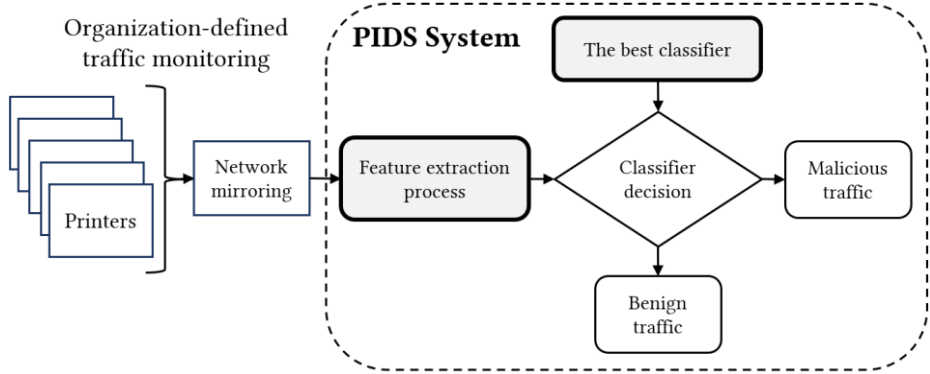

**Fig. 2.** PIDS architecture implementation

### 3.1 Data Collection

**Feature extraction.** The input of the system is the network traffic of the printers that are being monitored by the protection framework. The network traffic could be provided as real-time traffic or as recording network files in known formats like PCAP files. Real-time traffic could be captured in the network using implemented network monitoring devices (Network TAPs) [18,19] or as a simple configuration of printer port mirroring [20]. Our analysis of the printing protocols is based on analyzing the TCP sessions of the communication. Each traffic session (from SYN to FIN) will be analyzed by the framework. The results of this component will be an advanced vector of features that represent each session. To maintain reasonable performance, the system produces all the features from analyzing only the metadata of traffic.

Our focus in this research is merely on analyzing metadata information network traffic as opposed to deep payload analysis. The main reason for this decision is the performance constraint guiding us towards developing a light-weight IDS. This will enable simpler the deployment option of PIDS on the printer itself, as close as possible to the printer's motherboard and OS software. Another reason for choosing metadata analysis was the fact that every printer manufacturer may encode the inner payload of the printed job differently over the printing protocol. Therefore, PIDS with a payload inspection approach would require a different feature extraction module for each manufacturer.

There are three main categories of features we capture: size, time, and TCP communication properties. Each category of features has many advanced statistics features that we were able to calculate: averages, medians, stdev (standard deviations), variances, minimum, maximum, and ratios between the two sides of the connection. The two sides are named A and B, where A refers to the entity that initiated the TCP connection (sent the first SYN packet), and B refers to the entity that responded to the initiator A. The addition of this two-sided perspective increases the classifier's capacity to learn the benign and malicious behavior of the communicating sides. The behavioral model of each side of the network is different. Printers have their own unique behavior profile (like other communicating network entities which have their own behavior profiles) which we want to learn. The complete feature extraction process is depicted in Figure 3. The total number of features that are analyzed and extracted for each session is 75. This vector of features is built from 28 size features, 25 time features, and 22 TCP general features. A description of the three categories of features follows: (1) Size - examples are: the number of bytes each side of the connection sent, the size of the packets sent, the size ratio of the data that was sent and received by side A during the session, and the statistics calculation we described above. (2) Time - example for features are: total duration of the session, time interval between sending and receiving each packet, and statistics calculation on each side separately along with calculation on the all session together. (3) TCP properties - include features like: number of TCP flags appearances (ack, urg, push, reset), number of packets sent on the session, and as we described above also statistics calculations. The full list of features is provided in appendix A.

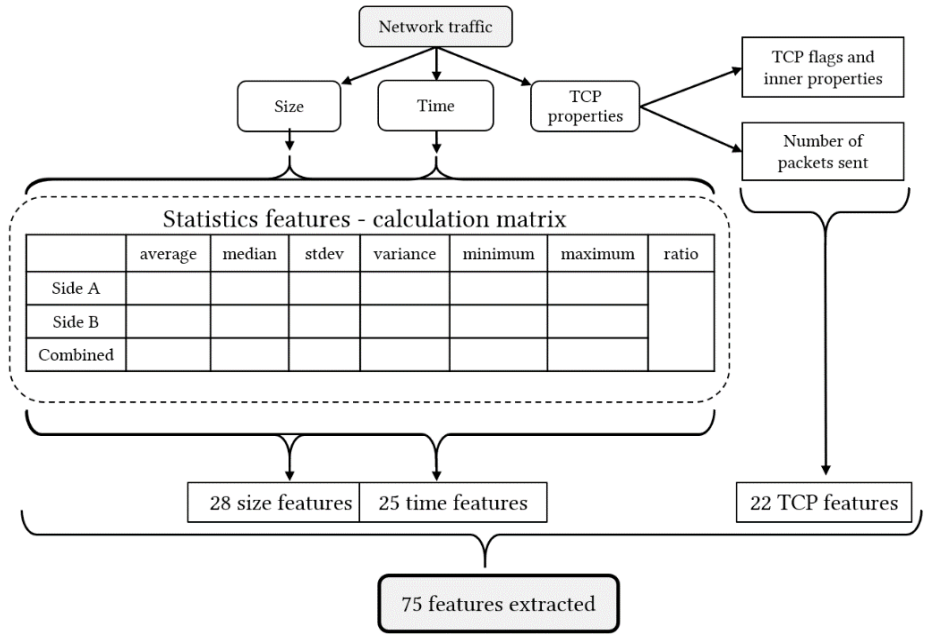

**Fig. 3.** Feature selection and generation.

**Dataset creation.** Generally speaking, one of the main challenges in research is how to achieve true, objective, and authentic datasets for experimentation. Our dataset contains thousands of malicious and benign labeled instances. Each instance is a feature vector we specifically designed and extracted from the printers' network traffic as described in 3.1. The instances were produced with our unique traffic data collection and generation component and were labeled separately as malicious and benign traffic sessions according to the following description.

*Benign Portion of the Dataset.* To collect valid benign printing protocol sessions, we recorded two months of traffic using network printers located in our labs' offices (including eight different printers), based on a reasonable assumption that the actual printers used in our labs' offices are not under attack during this period. This assumption is supported by the following facts: (1) In general, our offices' network is well secured by the official university security team, and it adheres to the latest security standards. (2) We performed the recording between June and July 2016 almost a year before the open-source printer attacking tool, PRET, was published. After the publication regarding PRET, printer attacks became much more feasible and easy to perform. Before PRET was announced, attacking printers was more complicated and performed by sophisticated APTs (advanced persistent threat), as opposed to "average level" attackers, script kiddies, and pen testers, and that is why we assume the printers in our offices network were functioning properly. Table 1 provides detailed information on the printing protocol sessions we acquired during this two months recording period. The total number of benign sessions obtained in this phase of the dataset building process is 8,213 benign sessions.

In addition, we purchased two new printers that were used solely for this research. We also recorded the benign printing protocol behavior of the new printers to make our benign dataset more robust. We printed 300 random documents on each of them, including different file formats: doc, docx, pdf, txt, ppt, pptx, and jpeg.

Therefore, the benign portion of the research dataset consists of a total of 8,813 trustworthy benign sessions.

**Table 1.** Benign sessions in the dataset.

| Printer model | Number of sessions initiated by the printer | Number of sessions initiated toward the printer | Total number of sessions |
|---|---|---|---|
| 2 months of real recoding of office printers: | | | |
| OKI 430DN | 16 | 2744 | 2760 |
| OKI 431DN | 17 | 355 | 372 |
| OKI 431DN | 17 | 224 | 241 |
| Xerox Phaser 8860 | 21 | 858 | 879 |
| Xerox Phaser 8860 | 2 | 39 | 41 |
| Xerox Phaser 8860 | 23 | 3098 | 3121 |
| Xerox 6605DN | 37 | 356 | 393 |
| Brother L2740DW | 20 | 386 | 406 |
| OKI 430DN | 16 | 2744 | 2760 |
| OKI 431DN | 17 | 355 | 372 |
| Self-initiated print jobs: | | | |
| HP OfficeJet Pro 8710 | 0 | 300 | 300 |
| OKI B432 | 0 | 300 | 300 |
| **A total number of 8,813 benign sessions were recorded.** | | | |

*Malicious Portion of the Dataset*. The purpose of the malicious portion of the dataset is to reflect real attack scenarios in various settings (Table 2). To collect such traffic, we recorded the network traffic while executing printing protocol attacks against different printers. We developed a special extension for PRET using the guides on the website, "www.hacking-printers.net" [21]. The extension was crafted especially for performing the attacks in an automated and completely randomized manner. To facilitate robust experiments, we needed many malicious sessions so that the classifier algorithm could learn the behavior of the attacks and their properties in the TCP session. Our PRET extension has two main capabilities - (1) Automation, and (2) randomization.

(1) *Automation* - the extension has the ability to execute different attacks in a predefined loop. There is a parameter that defines how many malicious sessions the tool should execute and forward to the target printer. We used this parameter to execute between 20 and 1,000 abnormal sessions on each printer. Each of those malicious sessions includes between one and 20 abnormal printing protocol commands. The extension has a bank of abnormal commands that are executed by the PRET framework, and these commands are listed in Table 2. An explanation of each of the commands can be found in the official GitHub page of the PRET tool [5].

**Table 2.** PRET commands divided by printing protocol and language type.

| General | PJL | PostScript | PCL |
|---|---|---|---|
| Ls | id | Id | info fonts |
| Get | version | version | info macros |
| Find | printenv | devices | info patterns |
| Cat | env | uptime | info symbols |
| Cd | nvram dump | date | info extended |
| Pwd | nvram read | pagecount | |
| chvol | info "xyz" | known | |
| Traversal | restart | search | |
| fuzz path | status | dicts | |
| fuzz blind | pagecount | resource | |
| Mirror | set | dump | |
| Df | display | restart | |
| Free | offline | overlay | |
| Put | reset | cross | |
| append | selftest | replace | |
| delete | flood | capture | |
| rename | lock | hold | |
| Edit | unlock | set | |
| touch | hold | lock | |
| mkdir | nvram write | unlock | |
| fuzz write | | reset | |
| | | config | |

Another automation capability we added is the option to attack different printers automatically. The tool gets a predefined bank of printers IPs, and PRET sent the commands to each of them in a continuous manner.

(2) *Randomization* - in order to create ensure a diverse pool of malicious instances, the randomization of the automated abnormal sessions was crucial. The extension implements a number of randomization levels to implement various attack scenarios in order to reflect the various strategies and behavior patterns of attackers. Figure 4 presents the randomization levels used by the extension. The importance of randomization stems from the fact that in reality attacks can exhibit a high degree of variance. Sometimes an attacker will choose to send malicious commands to the printer separately and use a different communication session for each sent command. While at other times he/she will attempt to execute a few commands in a cascade during the same communication session initiated with the target printer. Furthermore, a specific attacker may choose to start with a specific command while others may prefer a different execution order for their commands. In addition, the interval between sending the commands can be changed - some attackers will have automated tools and others will perform the attacks manually, and therefore we expect a large amount of variance in the time intervals between sending the commands in a specific attacking session.

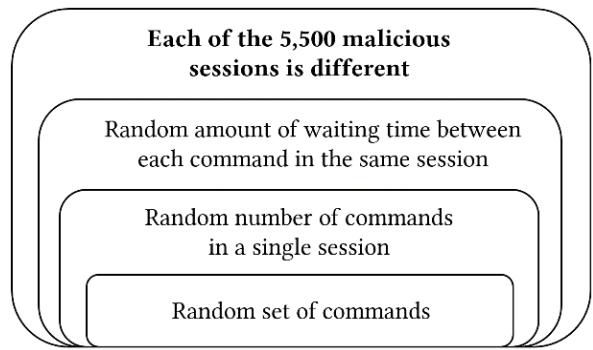

**Fig. 4.** Randomization of sessions.

Each of the commands in Table 2 is transformed to its corresponding low-level printing protocol command. Any other future printers attacking tool will still need to render and forward the same lower-level malicious printing protocols commands and thus our implementation is generic. This is also the reason why we configured PIDS as a network-based IDS network as opposed to a host-based IDS. Our system focuses only on the network behavior of such printing protocol attacks and their deviation from a benign behavior. Our low-level network protocol analysis is robust even when tools newer than PRET will emerge. Moreover, to increase reliability and accommodate a variety of attacks, we added randomization mechanisms to our attack generation tool based on PRET. Figure 5 describes the creation process of those malicious printing protocols' commands.

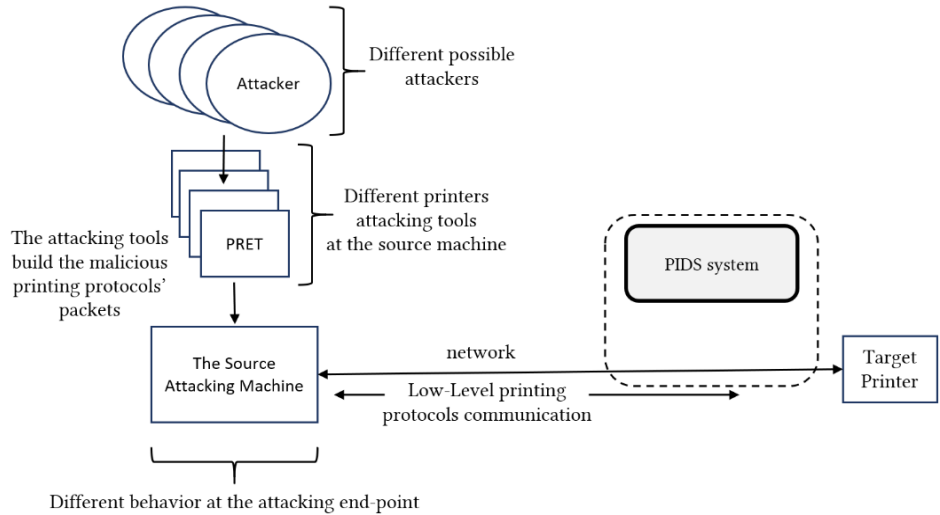

**Fig. 5.** The creation process of malicious printing protocols' commands

Table 3 summarizes the malicious printing protocols obtained when the attacks were executed with PRET.

**Table 3.** Malicious sessions in the dataset.

| # | Printer model | Command language | Number of sessions |
|---|---|---|---|
| 1 | OKI B432 | PJL | 1000 |
| 2 | OKI B432 | PCL | 200 |
| 3 | HP OfficeJet Pro 8710 | PCL | 1000 |
| 4 | HP OfficeJet Pro 6830 | PCL | 1000 |
| 5 | HP LaserJet 400 MFP | PJL | 1000 |
| 6 | HP LaserJet 400 MFP | PostScript | 1000 |
| 7 | HP LaserJet 400 MFP | PCL | 100 |
| 8 | HP LaserJet 600 M601 | PJL | 20 |
| 9 | HP LaserJet MFP M521dn | PJL | 20 |
| 10 | HP LaserJet MFP M426fdn | PJL | 20 |
| 11 | HP LaserJet MFP M507dn | PJL | 20 |
| 12 | HP LaserJet MFP M277dw | PJL | 20 |
| 13 | HP LaserJet MFP M277dw | PJL | 20 |
| 14 | HP LaserJet MFP M476dn | PJL | 20 |
| 15 | HP LaserJet 400 M401 | PJL | 20 |
| 16 | HP LaserJet 400 M401 | PJL | 20 |
| 17 | HP LaserJet P2055 | PJL | 20 |

**A total number of 5,500 malicious sessions were recorded.**

## 3.2 Training The Detection Model

The system evaluates each session of the printing protocol traffic (TCP connections from SYN to FIN), and each session with its special extracted feature vector is evaluated with the classifier. The classification decision making is an ongoing online process - as a session gets to the system for evaluation it will be immediately classified for real-time detection. An effective detection method should accurately distinguish between benign and malicious printing protocol network sessions.

In order to serve as a protective solution for users around the world it need to have outstanding detection rates. The experiments also focused on finding the highest scoring features, using the three different feature selection methods that will be described further in the results section. From our perspective, it is crucial that the system be practical and lightweight; therefore, we aimed to decrease the number of features that are used from the total number of 75 features to just 10 features.

We used five-fold cross-validation [22] to test and measure the detection performance. The dataset was divided into five subsets, and during each iteration, one of the five subsets was used as the test set and the other four subsets were merged to form the training set. Then the average results across all five trials was computed. Cross-validation (CV) minimizes the overfitting problem, because the training samples are independent from the validation samples. The popularity of the CV method largely comes from the randomization and heuristics of the data splitting. This evaluation technique provides robust accuracy in the experiment testing process.

The goal of the experiment was to validate the system's detection algorithm on the research. In the experiment we compared the different algorithms and their capabilities of detecting malicious printing protocol commands. In addition, we performed feature selection using various methods to locate the most powerful features for distinguishing between benign and malicious traffic.

## 3.3 Evaluating The Detection Model

In this section we present the results of our experiment and provide the detailed detection results. All of the instances in our dataset (as explained in section 3.1) were used in this experiment.

**Testing of Classification Algorithms**. A comparison of the results of all five tested algorithms is provided in Table 4, while Figure 6 focuses on presenting the accuracy measurements.

**Table 4.** Comparison of the five classification algorithms' results.

| Algorithm | FPR | TPR | AUC | Accuracy | Length of time it takes to build the model (in seconds) |
|---|---|---|---|---|---|
| K-Means | 0.313 | 0.761 | 0.721 | 75.46% | 0.30 |
| Decision Tree C4.5 | 0 | 1 | 0.999 | 99.95% | 0.54 |
| Naïve Bayes | 0.311 | 0.507 | 0.816 | 50.74% | 0.19 |
| Bayesian Networks | 0.005 | 0.995 | 1 | 99.55% | 0.57 |
| SVM | 0.016 | 0.987 | 0.986 | 98.69% | 1.72 |

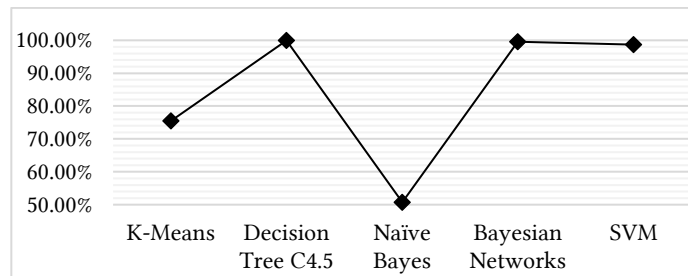

**Fig. 6.** Accuracy results for the five tested algorithms.

As we wanted our PIDS protection system to be very practical and feasible for actual implementation (even on the printer motherboard), PIDS should be light-weight and very effective in its performance. Therefore, we chose to focus and examine PIDS's detection rates with the above 5 basic algorithms and not with more complicated machine learning algorithm, such random forest or deep learning. We observed outstanding results with the selected algorithms.

We evaluated the algorithms' performance using the following metrics: FPR (False Positive Rate), TPR (True Positive Rate), AUC (Area Under ROC Curve) and total Accuracy. We observe that three of the algorithms (namely, the Decision Tree, Bayesian Networks, and SVM) provided very good results. The Decision Tree algorithm performed the best in terms of all of the metrics. The confusion matrix for the Decision Tree classifier is depicted in Table 5.

**Table 5.** Confusion matrix for the Decision Tree classifier.

| Classified as | a | b |
|---|---|---|
| a = benign | 8809 | 4 |
| b = malicious | 2 | 5498 |

It is also interesting to look at the decision tree itself and the features that were used. The constructed decision tree is provided in Figure 7. The algorithm mainly used a few interesting features to correctly classify the instances (see appendix B for a description of these features).

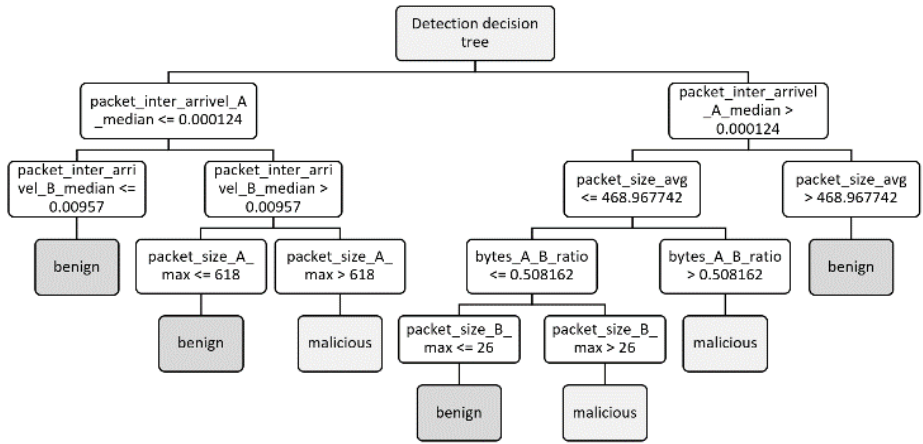

**Fig. 7.** The decision tree classifier's output.

From our results we can see that the malicious use of printing protocols is associated with a longer time interval between the communication packets. One possible reason for this is that when malicious commands encapsulated in the printing protocols are sent, the receiving printer analyzes the commands and then sends back the results of those abnormal commands - a process which results in a longer time interval between each packet that is sent by the printer. For example, for benign commands like those typically used for printing jobs, the printer just needs to send back a quick response that the printed job was indeed received, however a malicious command to read from the printer's memory will cause the response time from the printer to be slower.

Another interesting observation is that in 98% of the recorded sessions the printer was on side B and not side A. It means that most of the times the printers were not the side that initiated the connection. In those scenarios we analyzed the bytes_A_B_ratio feature, and found that about 70.45% of all of the benign printing commands were executed at a size ratio significantly lower than 0.38. This means that the network entity that communicates with the printer sends almost twice as much data to the printer than the amount of data sends by the printer. For example, when sending a print job of 500 kilobytes, typically the printer just sends back a few packets of acknowledgments totaling less than 190 kilobytes. On the other hand, in a possible attack scenario the attacker sends a very short command and might get an equal amount of data, or even more data, from the printer in return. As a case in point, an abnormal reconnaissance command of querying the target printer for its type would include sending 528 bytes of data to the printer and receiving back 485 bytes of data (including the printer's model name). Another more drastic contrast occurs when sending a command to read a print job from the printer's memory; this would entail sending a very short message (the malicious reading command) and then receiving the full print job which is obviously larger than the initial read command.

In addition, when examining the packet_size_B_max feature we note that in 98.67% of all of the benign sessions, the maximal data transfer in a single packet from the printer is less than 50 bytes. This is because in a benign session the printer usually does not send much data back to the client entity, as most of the responded packets are merely acknowledgments and small predefined bit flags for updates about the printer's status (when the printer is busy or has finished printing the print job, when the printer is out of paper, etc.). In malicious sessions, however, only 9.04% of the sessions had a packet_size_B_max of less than 50 bytes, since the responses in such scenarios often contain leaked data to the attacker.

The Naïve Bayes and K-Means algorithms had a 30% false positive rate and therefore cannot be applied in a practical PIDS. Bayesian Networks and SVM performed well (achieving over 98.5% accuracy) but also had a higher false positive rate than Decision Tree algorithm. Moreover, training using SVM takes three times longer than training using the Decision Tree and Bayesian Networks; therefore, SVM was not selected for our suggested system. Another point to mention is the performance consideration when choosing to perform the training stage on large corporate networks comprised of many printers. Advanced incremental training can be used to address this problem as demonstrated by the active learning approach [23]. Therefore, the selected algorithms for our evaluation were Decision Tree C4.5 and Bayesian Networks.

**Feature Selection.** In order to create a fast, lightweight PIDS, we wanted to identify the 10 most valuable features from all the 75 features we collected. We compared the following feature selection methods that were designed to select the top features: InfoGain [24], GainRatio [25], and correlation (Pearson) [26].

These three methods were applied, and the features with the best scores are presented in Figure 8, along with their exact given score.

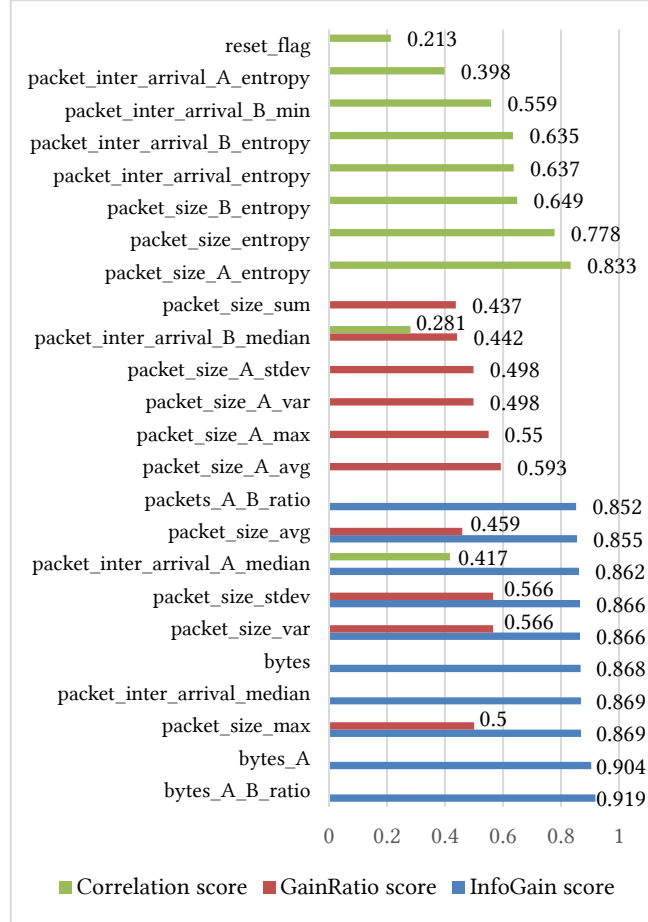

**Fig. 8.** Top scored features.

After completing the abovementioned phase of the feature selection process, the next step is to evaluate the detection rates of the framework when it uses only the 10 top scoring features. The accuracy results for this scenario are listed in Table 6.

**Table 6.** Experimental results when using only the top 10 features from the different three selection methods.

| Feature selection method | Algorithm | Accuracy |
|---|---|---|
| InfoGain | Decision Tree C4.5 | 99.94% |
|  | Bayesian Networks | 99.77% |
| GainRatio | Decision Tree C4.5 | 99.92% |
|  | Bayesian Networks | 98.69% |
| Correlation | Decision Tree C4.5 | 99.88% |
|  | Bayesian Networks | 98% |

Like the evaluation performed of the 75 features, this evaluation was based on five-fold cross-validation and performed on the same dataset. We compared the algorithms that performed best in first phase of the experiment: Decision Tree and Bayesian Networks. As can be seen in the table, we still obtain high detection scores when the framework only focuses on analyzing 10 features; in addition, in each case, the Decision Tree performs slightly better than the Bayesian Networks.

In this process of feature selection, we also found that there were six features which were selected to be among the top 10 features of two selection methods. The fact they were chosen by more than one method indicates their importance in the detection process. These leading features are: packet_size_var, packet_size_stdev, packet_size_max, packet_size_avg, packet_inter_arrival_A_median, packet_inter_arrival_B_median (appendix B). Four of them are variants of packet size calculation which indicates that size has a significant impact on the classifier's ability to differentiate between the malicious and benign use of printing protocols. Two other features calculate the median of the time duration between the arrival of packets, one of them referring to the arrival times on side A, and the other refers to the same thing for side B. The reason for this could be that in the case of attack commands the printer needs more processing time in order to respond back to the attack side, and therefore the time interval between sent packets is larger. From the attacker network side, the time interval between sending multiple malicious commands could be higher than the time interval required for the automatic processing and sending of documents. Another important observation that can be made is that the 10 top scoring features overall have very high accuracy scores, and a detection classifier can be built from just these features (and even from just the top six features. This is because of the high score of the features (for example, the fact that the size features obtained a score of over 0.85 using the InfoGain method and about 0.5 using the GainRatio scale). This conclusion can be reinforced again by analyzing the decision tree presented earlier that was built from 12 features and provides excellent detection and performance rates.

# 4     Discussion

The important research conducted in [12] on attacking printers revealed new threats against the printers used daily by individuals and organizations around the world and pointed to the need to improve the security mechanisms of printers. Follow our experiment we can observe that attacking commands from the open-source attacking tool, PRET, (the most popular attacking tool used against printers) could be detected by our framework. The practical implementation of our proposed framework could be achieved using an external analysis system or on the printers themselves, where each printer would be equipped with its own attacking detection component. The printer's firmware could analyze the aforementioned protocols' metadata traffic features to detect and abort abnormal printing protocols and issue anomaly alerts based on this detection. Another option is to use the suggested detection framework with passive learning of the actual monitored printers and then train and update the detection classifier incrementally in real-time. After a short learning period the detection framework would be ready to use and start issuing alerts regarding abnormal malicious printing protocol usage with high accuracy and a near zero false positive rate as indicated by our research.

Moreover, PIDS is also designed to detect an already compromised printer and not only the preliminary printer infection stage of an attack. PIDS would alert on network anomalies, including the common attack scenarios of a printer that sends copies of its users' printed jobs to an external target IP. It would also alert about differences in the size ratio between the data sent/received by the printer to/from an external target IP (as opposed of receiving more data in a benign scenario, the printer will send out more data when infected). Another scenario could be an attacker that compromised a printer to serve as a launchpad for executing malicious code. The attacker can perform malicious actions like network scanning of peer machines or operate stealthy C&C communications to other already compromised machines on the attacked network. Such malicious network communications would be different than thousands of real-life benign printing sessions we have researched and were used to train the PIDS's classifier.

An interesting comparison can be made between our research and [14]. While both of the studies deal with different types of threats, their research aims to detect physical damage in additive manufacturing (CNC milling) caused by cyber-attacks, and ours aims to detect attacks on network printers. Both of the studies used ML algorithms for the detection process. However, this research achieved 99.9% accuracy using a simple C4.5 Decision Tree, while in their research they achieved 91.1% accuracy with the advanced Random Forest algorithm. A possible explanation for this difference is that their dataset is more complicated than ours. It will be interesting to evaluate the use of our framework in an additive manufacturing environment, using the 75 features extracted from the associated network traffic and benchmark our performance again, so we can get a sense of whether our framework is also effective in the protection of 3D printing and CNC milling which have been shown vulnerable to cyber-attacks.

## 5   Conclusions

This research demonstrates the proposed system's ability to improve the security of printers. Our results suggest that a supervised ML detection framework can provide nearly absolute detection rates for identifying malicious printing protocols traffic.

The limitation of our research lies in the sole use of a supervised algorithm. We evaluated the detection framework on our own malicious printing protocol usage, however in the future there might be new types of attack scenarios and printer attack tools that will challenge the performance of our classifier (because the classifier wasn't trained on them). Our future work will focus on improving the detection framework and addressing the limitation mentioned above by using unsupervised ML algorithms, and more specifically, anomaly detection algorithms; for example, we plan to adapt the framework so it builds network behavior profiles of the monitored printers in order to detect anomalies from this learned profile. In such a method there is no need to learn malicious behavior profiles in advance, because the framework will issue an anomaly alert when a deviation from the proper printing protocol (normal daily printer's behavior) is detected. Another line of work will increase the scope of the framework so it can be used to protect other devices such as 3D printers, IP phones, smart IP projectors, etc. We will need to research the specific operating protocols for each type of device, to record and learned their benign associated network traffic, perform multiple attacks against them, and record the malicious usage of their unique protocols. After compiling thoroughly and trustworthy datasets, the learning algorithm could be used to create a more general robust detection classifier for those new types of devices.

# Appendix A: List Of Extracted Features

**Table 7.** List of all extracted features categorized by type

| # | Feature Name | Feature Type | # | Feature Name | Feature Type |
|---|---|---|---|---|---|
| 1 | ack | TCP Properties | 39 | packet_size_A_min | Size |
| 2 | ack_A | TCP Properties | 40 | packet_size_A_stdev | Size |
| 3 | ack_B | TCP Properties | 41 | packet_size_A_sum | Size |
| 4 | bytes | Size | 42 | packet_size_A_var | Size |
| 5 | bytes_A | Size | 43 | packet_size_B_avg | Size |
| 6 | bytes_A_B_ratio | Size | 44 | packet_size_B_entropy | Size |
| 7 | bytes_B | Size | 45 | packet_size_B_max | Size |
| 8 | ds_field_A | TCP Properties | 46 | packet_size_B_median | Size |
| 9 | ds_field_B | TCP Properties | 47 | packet_size_B_min | Size |
| 10 | duration | Time | 48 | packet_size_B_stdev | Size |
| 11 | packet_inter_arrival_A_avg | Time | 49 | packet_size_B_sum | Size |
| 12 | packet_inter_arrival_A_entropy | Time | 50 | packet_size_B_var | Size |
| 13 | packet_inter_arrival_A_max | Time | 51 | packet_size_avg | Size |
| 14 | packet_inter_arrival_A_median | Time | 52 | packet_size_entropy | Size |
| 15 | packet_inter_arrival_A_min | Time | 53 | packet_size_max | Size |
| 16 | packet_inter_arrival_A_stdev | Time | 54 | packet_size_median | Size |
| 17 | packet_inter_arrival_A_sum | Time | 55 | packet_size_min | Size |
| 18 | packet_inter_arrival_A_var | Time | 56 | packet_size_stdev | Size |
| 19 | packet_inter_arrival_B_avg | Time | 57 | packet_size_sum | Size |
| 20 | packet_inter_arrival_B_entropy | Time | 58 | packet_size_var | Size |
| 21 | packet_inter_arrival_B_max | Time | 59 | packets | TCP Properties |
| 22 | packet_inter_arrival_B_median | Time | 60 | packets_A | TCP Properties |
| 23 | packet_inter_arrival_B_min | Time | 61 | packets_A_B_ratio | TCP Properties |
| 24 | packet_inter_arrival_B_stdev | Time | 62 | packets_B | TCP Properties |
| 25 | packet_inter_arrival_B_sum | Time | 63 | push | TCP Properties |
| 26 | packet_inter_arrival_B_var | Time | 64 | push_A | TCP Properties |
| 27 | packet_inter_arrival_avg | Time | 65 | push_B | TCP Properties |
| 28 | packet_inter_arrival_entropy | Time | 66 | reset | TCP Properties |
| 29 | packet_inter_arrival_max | Time | 67 | reset_A | TCP Properties |
| 30 | packet_inter_arrival_median | Time | 68 | reset_B | TCP Properties |
| 31 | packet_inter_arrival_min | Time | 69 | tcp_analysis_duplicate_ack | TCP Properties |
| 32 | packet_inter_arrival_stdev | Time | 70 | tcp_analysis_keep_alive | TCP Properties |
| 33 | packet_inter_arrival_sum | Time | 71 | tcp_analysis_lost_segment | TCP Properties |
| 34 | packet_inter_arrival_var | Time | 72 | tcp_analysis_out_of_order | TCP Properties |
| 35 | packet_size_A_avg | Size | 73 | urg | TCP Properties |
| 36 | packet_size_A_entropy | Size | 74 | urg_A | TCP Properties |
| 37 | packet_size_A_max | Size | 75 | urg_B | TCP Properties |
| 38 | packet_size_A_median | Size | | | |

## Appendix B: Leading Features

**Table 8.** Explanation of leading features

| Feature Name | Feature Explanation |
| --- | --- |
| packet_size_var | Variance of the packets' sizes |
| packet_size_stdev | Standard deviation of the packets' sizes. |
| packet_size_max | Maximal size of all of the packets. |
| Packet_size_avg | Average size of all of the packets. |
| Packet_inter_arrival_A_median | Median of the time duration between the arrival of packets to side A. |
| Packet_inter_arrival_B_median | Median of the time duration between the arrival of packets to side B. |
| Packet_size_A_max | Maximal size of data sent by side A in a single packet. |
| packet_size_B_max | Maximal size of data sent by side B in a single packet. |
| Packet_A_B_ratio | Ratio of sizes (ratio of the traffic sizes between the data side A received and the data size A sent in the connection. |